# The Field Perturbation Theory of the Double Correlated Phase in HTSC

Moshe Dayan

Department of Physics, Ben-Gurion University,

Beer-Sheva, 84105, Israel.

e-mail: mdayan@bgu.ac.il

# The Field Perturbation Theory of the Double Correlated Phase in HTSC

Moshe Dayan
Department of Physics, Ben-Gurion University,
Beer-Sheva, 84105, Israel.

Abstract

The double-correlated phase in HTSC, and its treatment by field perturbation theory, is being established. In particular, we define the ground state, the quasi-particle excitations, and construct an appropriate field. We also derive the unperturbed Hamiltonian, and the propagators for the unperturbed state. Then we discuss the perturbation Hamiltonian, and show that the Hartree Diagram is significant for both the pseudogap and the superconductive order parameters, and suggest that it yields the major contribution to these parameters.

1. **INTRODUCTION**

In a few former papers, I suggested a field perturbation theory for both the superconductive and the pseudogap order parameters in HTSC [1-3]. The theory is based on the field theory of Nambu, and Gorkov [4], for the ordinary superconductive problem. To generalize the theory to HTSC, in which electron correlations exist also in the normal state, one has to assume double correlations, in order to incorporate the correlations of the superconductive Cooper-pairs, with the correlations of the pseudogap electron-hole pairs. This was done in [2], by enlarging the field dimensionality to four. The double correlations model assumes correlations, not only between the Cooper-pair quasi-particles $(\mathbf{k}, s : -\mathbf{k}, -s)$, and between their associates $(\bar{\mathbf{k}}, s : -\bar{\mathbf{k}}, -s)$, where $|\mathbf{k} - \bar{\mathbf{k}}| = 2k_F$, but also assumes that the latter are correlated with the former to produce the pseudogap. The inherent physical source of the particle-hole pairing is the Fermi surface nesting, which is known to causes the breakdown of the normal metallic state [5-7]. Fermi surface nesting was indeed proposed to exist in HTSC by several investigations [8-11]. The particle-hole pairing model, which was proposed to resolve the nesting driven singularities, yields internal CDW or SDW that violate momentum conservation by $2K_F$, in directions normal to the nesting surfaces. These field waves have been observed directly [12,13]. Their consequential violation of momentum conservation by $2K_F$ is well demonstrated by the "shadow Fermi surfaces" that have usually been observed in AREPS [14,15].

The first and most impressive experimental feature, which any theory (of HTSC) should account for, is the large values of the order parameters- the superconductive, and even more so- the pseudogap. Limitations imposed by lattice stability requirements, and by fundamental features of the dielectric function [16], prohibit the Fock integral from being the sole source of these large order parameters. Former works [1-3] proposed that Hartree integrals are the main sources of the large order parameters. To enable such proposals, within the matrix propagator formalism, one has to assume that interaction vertices are spanned by the same matrices that span the self-energies. In [1,2], such assumptions were conjectured mainly on intuitive basis. A more careful examination might question the validity of such vertices. This is so because in the Nambu formalism, the superconductive order parameter is given by the $\tau_1$ Pauli matrix, while the interaction vertex is given by the $\tau_3$ Pauli matrix. The correlation pseudogap in [2] is also given by the $\tau_1$ Pauli matrix, while the interaction vertex is presented by the identity matrix. However, while the interaction vertices are diagonal when presented by the uncondensed state, they may acquire non-diagonal components when expressed by the fields of the condensed states. This was proved for superconductivity as well as for the correlation pseudogap [3, 17]. Moreover, the internal CDW or SDW fields (of $2K_F$ wave-numbers) may scatter quasi-particles and break their momentum definiteness. Thus, in addition to the momentum defined propagator $G_0(k,\omega)$, one has the momentum undefined propagators $G_0(k,\bar{k}\omega)$, and $G_0(\bar{k},k\omega)$, where $\bar{\mathbf{k}} = \mathbf{k} - (\mathbf{k}_\perp/|\mathbf{k}_\perp|) \cdot 2k_F$. This, in turn, facilitates the introduction of the off-diagonal vertex- $\tau_1$, and consequently, the Hartree diagram for the order parameter.

For ordinary superconductivity, off-diagonal vertices are prohibited for the known Nambu field. However, it was shown that, with the definition of the proper coherence field, quasi-particles may also be scattered via an off-diagonal vertex [17]. It was also shown that, although the off-diagonal Hartree diagram is valid in principle, it yields zero for ordinary superconductors, due to the vanishing total interaction at zero energy and momentum.

When comparing between the discussed problems, the ordinary superconductivity, and the pseudogap problem in the normal state, we notice the following difference. In the first case, the Nambu field does not yield off-diagonal vertices, and the introduction of the coherence field is essential for it. In the second case, the equivalent to the Nambu field is sufficient for the obtainment of the off-diagonal vertex. However, it has been shown, that coherence fields may also be defined for the pseudogap problem in the normal state [3]. The Nambu-like field is defined by

$$\tilde{\Psi}_{k,s} = \begin{pmatrix} -u_k \\ v_k \end{pmatrix} \gamma_{k,s} + \begin{pmatrix} v_k \\ u_k \end{pmatrix} \eta^+_{\bar{k},s} \,. \tag{1}$$

The coherence field is defined by

$$\Psi_{k,s} = \begin{pmatrix} -u_k \\ v_k \end{pmatrix} \gamma_{k,s} - \begin{pmatrix} v_k \\ u_k \end{pmatrix} \eta^+_{\bar{k},s}, \tag{2}$$

with $\gamma_{k,s} = -u_k c_{k,s} + v_k c_{\bar{k},s}$, and $\eta^+_{k,s} = v_k c_{k,s} + u_k c_{\bar{k},s}$. The quantities $v_k$ and $u_k$ are the probability amplitudes for having the state k full or empty in the condensed state, respectively. Both the above fields are compatible with the same unperturbed Hamiltonian and unperturbed propagator. However, they are not compatible with the same interaction vertices. While for the Nambu-like field the interaction vertex for the Hartree diagram is

$$\tilde{I}^H_{k,s} = \tilde{\Psi}^+_{k',s} \{\delta_{k,k'} I + \delta_{k,\bar{k}'} \tau_1\} \tilde{\Psi}_{k,s}, \tag{3a}$$

for the coherence field it is

$$I_{k,s}^H = \Psi_{k',s}^+ \{I\delta_{k,k'} + \delta_{k,\bar{k}'}[(-1+8u_k^2 v_k^2)\tau_1 - 4u_k v_k(u_k^2 - v_k^2)\tau_3]\}\Psi_{k,s} \ . \tag{3b}$$

With the relations $\varepsilon_k = E_k(u_k^2 - v_k^2)$, and $\Lambda_k = -E_k(2u_k v_k)$, we get

$$I_{k,s}^H = \Psi_{k',s}^+ \{I\delta_{k,k'} + \delta_{k,\bar{k}'}E_k^{-2}[-(\varepsilon_k^2 - \Lambda_k^2)\tau_1 + 2\Lambda_k \varepsilon_k \tau_3]\}\Psi_{k,s} \tag{3c}$$

Therefore, for the coherence field, we have an additional $\tau_3$ vertex, and the $\tau_1$ and $\tau_3$ vertices are weighted by coherence factors. The additional $\tau_3$ vertex enables an Hartree self energy correction to the uncondensed excitation energies $\varepsilon_k$, transforming it into $\hat{\varepsilon}_k = \varepsilon_k + \chi_k$. A tentative evaluation of the ground state energy $[-\sum_{k<k_F,s}\sqrt{\hat{\varepsilon}_k^2 + \Lambda_k^2}]$ shows that, due to $\chi_k$, it is lower for the coherence field than for the Nambu-like field, making the first a better choice over the second.

The present paper extends the discussed analysis to the double correlation phase by defining the double correlated ground state, the excitations, the field, the unperturbed propagator, and the unperturbed Hamiltonian. Then it derives the perturbation Hamiltonian, and discusses the self-energy of the perturbed state. We put our attention mainly on Hartree diagrams, since they seem to be larger than their Fock counterparts, and also easier to evaluate. It has been shown that, contrary to ordinary superconductivity, the Hartree diagram for the pseudogap does not vanish, but yields a finite and significant contribution [3]. The present analysis proves that this contribution is enhanced by the double correlation phase. It also proves the significant contribution of the Hartree diagram to the superconductive gap in HTSC.

For the sake of convenience we assume zero temperature. We also disregard disorder for the same reason, although it is essential for the metallic and superconductive character of HTSC. The addition of disorder to the treatment, along the lines of [1], is tentatively discussed in the last section.

## 2. THE DOUBLE CORRELATED GROUND STATE AND ITS EXCITATIONS.

The ground state $|\Phi_0>$ involves correlations between pairs of the components of the four dimensional Nambu-like field

$$\tilde{\Psi}_k = \begin{pmatrix} c_{k,s} \\ c^+_{-\bar{k},-s} \\ c_{\bar{k},s} \\ c^+_{-k,-s} \end{pmatrix}. \tag{4}$$

The ground state of the double correlation state was defined in [2] as

$$|\Phi_0> = \prod_{\varepsilon_K<0}[v_k + u_k\tilde{\Psi}^+_k\alpha_1\tilde{\Psi}_k + w_k\tilde{\Psi}^+_k\alpha_3\tilde{\Psi}_k + \theta_k c^+_{\bar{k},s}c_{k,s}c^+_{-\bar{k},-s}c_{-k,-s}]|0> \tag{5}$$

where $|0>$ is the ground state of the uncondensed (free electron) phase. Notice that the pre-factors $u_k$, and $w_k$ in Eq. (4), are different than their counterparts in [2]. We also exchanged between $v_k$ and $u_k$, in accordance with [3]. The last term of Eq. (5) is necessary to make $|\Phi_0>$ an eigenstate of the unperturbed Hamiltonian. By demanding the normalization of the ground state, $<\Phi_0|\Phi_0>=1$, and assuming real coefficients, we get

$$v_k^2 + 2u_k^2 + 2w_k^2 + \theta_k^2 = 1. \tag{6}$$

Both the theory of superconductivity and the theory of correlation pseudogap suggest that the ground state should be either symmetric or anti-symmetric with respect to the transformation $\mathbf{k} \leftrightarrow \bar{\mathbf{k}}$. This is obtained by demanding that $|v_{\bar{k},k}| = |\theta_{k,\bar{k}}|$, $|u_{\bar{k}}| = |u_k|$, and $|w_{\bar{k}}| = |w_k|$. We assume positive signs for convenience. With such coefficients, the symmetry becomes more apparent when $|0>$ is written explicitly by means of the $c^+$-creation operators acting on the vacuum state.

We proceed in the same manner as in [3], and in [17]. We base our definitions of the proper fields on the elementary excitations of the system, as is usually done in field theories. We assume that the excitations are linear combinations of the four components of the vector field in Eq. (4). One such excitation is defined as

$$\gamma_k = a_k c_{k,s} + b_k c^+_{-\bar{k},-s} + r_k c_{\bar{k},s} + d_k c^+_{-k,-s} \tag{7}$$

To determine the coefficients, we apply the following requirement

$$\gamma_k|\Phi_0> = 0, \tag{8}$$

which ensures that $\gamma_k$ is indeed a destruction operator of a true elementary excitation. Eq. (8) translates into a linear combination of the four operators $c_{k,s}$, $c^+_{-\bar{k},-s}$, $c^+_{-\bar{k},-s} c_{-k,-s} c_{k,s}$, and $c_{k,s} c^+_{\bar{k},s} c^+_{-\bar{k},-s}$. Since each pre-factor must equal zero separately, we get four homogeneous equations for the coefficients in Eq. (7). After some standard operations, the equations reduce to the following

$$a_k u_k - b_k w_k + d_k \theta_k = 0, \qquad (9a)$$

$$r_k u_k + b_k v_k - d_k w_k = 0, \qquad (9b)$$

$$\theta_k v_k = u_k^2 + w_k^2. \qquad (10)$$

Eq. (6) and Eq. (10) yield

$$\theta_k + v_k = \pm 1. \qquad (11)$$

We assume the plus sign for convenience.

Eqs. (9) leave the coefficients of Eq. (7) undetermined unless we impose additional relations. One is a normalization relation. The other might be based on demanding that $\gamma_k^+ | \Phi_0 >$ is an eigenstate of the unperturbed Hamiltonian, $H_0 \gamma_k^+ | \Phi_0 >= E_k \gamma_k^+ | \Phi_0 >$. The problem is that such a relation yields a set of insoluble non-linear equations. One has to resort to intuition and impose some arbitrary relationship between the coefficients of Eq. (7). Each such relation produces only one set of coefficients, namely, one excitation for every **k**. Since there are two independent Bogoliubov-Valatin excitations in regular superconductors, and two independent excitations in HTSC in the normal state, we expect four independent excitations for each **k** in the double correlated state. Two of these excitations are quasi-particles, and the other two are quasi-antiparticles. Finding such a set involves a lot of work, because each set has to pass several time consuming tests. Each member of the set must be an eigenstate of the unperturbed Hamiltonian $H_0$, with the same eigenvalue $E_k$. A set which fails the last criterion does not produce a propagator with the desired form, one that is suitable for field perturbation theory. It turns out that Eqs. (7-9) are insufficient for obtaining the whole set of excitations. One has to add an equation, which is equivalent to (7), but with the conjugated operators of those in Eq. (4). This equation is

$$\rho_k = e_k c_{-k,-s} + f_k c^+_{\bar{k},s} + g_k c_{-\bar{k},-s} + h_k c^+_{k,s} \qquad (12)$$

Applying the conditions $\rho_k | \Phi_0 >= 0$ yields

$$-f_k v_k + g_k u_k + h_k w_k = 0, \qquad (13a)$$

$$f_k w_k + e_k u_k - h_k \theta_k = 0. \tag{13b}$$

One quasi-particle may be found from Eqs. (9) by assuming $a_k = r_k$, which yields

$$\hat{\gamma}_k = \hat{R}_k^{-1}[-u_k c_{k,s} + (\theta_k + w_k)c^+_{-\bar{k},-s} - u_k c_{\bar{k},s} + (v_k + w_k)c^+_{-k,-s}], \tag{14}$$

where $\hat{R}_k^{-1}$ is a normalization factor. The other quasi-particle is obtained from the same equations by assuming $b_k = -d_k$,

$$\hat{\eta}_k = \hat{R}_k^{-1}[(\theta_k + w_k)c_{k,s} + u_k c^+_{-\bar{k},-s} - (v_k + w_k)c_{\bar{k},s} - u_k c^+_{-k,-s}]. \tag{15}$$

One could also assume $a_k = -r_k$, and $b_k = d_k$. These assumptions yield equivalent operators that are not normal to $\hat{\gamma}_k$ and $\hat{\eta}_k$. Applying the same procedure on Eqs. (13) yields

$$\hat{\rho}_k = \hat{R}_k^{-1}[u_k c_{-k,-s} + (\theta_k + w_k)c^+_{\bar{k},s} + u_k c_{-\bar{k},-s} + (v_k + w_k)c^+_{k,s}], \tag{16}$$

$$\hat{\sigma}_k = \hat{R}_k^{-1}[(\theta_k + w_k)c_{-k,-s} - u_k c^+_{\bar{k},s} - (v_k + w_k)c_{-\bar{k},-s} + u_k c^+_{k,s}]. \tag{17}$$

The operators given by Eqs. (14-17) define a set of independent excitations that could construct a field, and a perturbation theory. However, when such a procedure is attempted, one finds inconsistency with the assumption that all the four excitations have the same eigenvalue. This was also the situation in the two dimensional problem of the pseudogap in the normal state [3]. There, however, the problem could be fixed by shifting the Fermi level. In the present problem, a single shift does not resolve the problem, because the common eigenvalue of two excitations is shifted to one direction while the common eigenvalue of the other excitations is shifted to the opposite direction. Moreover, the shifts are too large to be ignored. This problem has been a major difficulty and most time consuming, until the following solution is found.

A better set of excitations is obtained by the following combinations $\gamma_k = \hat{\gamma}_k + \hat{\eta}_k$, $\eta_k = \hat{\eta}_k - \hat{\gamma}_k$, $\rho_k = \hat{\rho}_k + \hat{\sigma}_k$, and $\sigma_k = \hat{\rho}_k - \hat{\sigma}_k$. The new set may be expressed in a matrix form

$$O_k = \begin{pmatrix} \gamma_k \\ \eta_k \\ \rho_k^+ \\ \sigma_k^+ \end{pmatrix} = P_k \tilde{\Psi}_k \tag{18}$$

where

$$P_k = R_k^{-1} \begin{pmatrix} \theta_k + w_k - u_k & \theta_k + w_k + u_k & -v_k - w_k - u_k & v_k + w_k - u_k \\ \theta_k + w_k + u_k & -\theta_k - w_k + u_k & -v_k - w_k + u_k & -v_k - w_k - u_k \\ v_k + w_k + u_k & -v_k - w_k + u_k & \theta_k + w_k - u_k & \theta_k + w_k + u_k \\ v_k + w_k - u_k & v_k + w_k + u_k & \theta_k + w_k + u_k & -\theta_k - w_k + u_k \end{pmatrix} \tag{19}$$

$$= R_k^{-1}[(\theta_k + w_k - u_k)\tau_3 + (\theta_k + w_k + u_k)\tau_1 + (v_k + w_k + u_k)\alpha_0\beta + (v_k + w_k - u_k)i\beta\alpha_2].$$

In Eq. (19), $R_k = \sqrt{2(1+2w_k)}$, $\alpha_2$ and $\beta$ are the known Dirac matrices, $\tau_1$ and $\tau_3$ are 4-dimensional Pauli matrices in which the 2-dimensional Pauli matrices are repeated along the diagonal of $P_k$. The matrix $\alpha_0$ is usually denoted, in the relativistic field theory, as - $\gamma^5$, and it contains two 2- dimensional unity matrices along the cross diagonal. Note that $P_k$ is orthogonal, $P_k^{-1} = P_k^+$. We have already commented that the symmetry properties of the ground state suggest that

$$u_{\bar{k}} = u_k, \tag{20a}$$

$$w_{\bar{k}} = w_k \tag{20b}$$

$$v_{\bar{k}} = \theta_k, \text{ and } \theta_{\bar{k}} = v_k. \tag{20c}$$

These last relations, together with the orthogonality of $P_k$, and the relations $\{\tilde{\Psi}_k^+, \tilde{\Psi}_k\} = 1$, and $P_{\bar{k}} = \alpha_1 P_k \beta$, suggest the following anti-commutation relations

$$\{\gamma_{k'}, \gamma_k^+\} = \{\eta_{k'}, \eta_k^+\} = \{\rho_{k'}, \rho_k^+\} = \{\sigma_{k'}, \sigma_k^+\} = \delta_{k,k'} \tag{21a}$$

$$\{\gamma_{k'}, \eta_k^+\} = \{\eta_{k'}, \gamma_k^+\} = -\delta_{\bar{k},k'} \tag{21b}$$

$$\{\sigma_{k'}, \rho_k^+\} = \{\rho_{k'}, \sigma_k^+\} = \delta_{\bar{k},k'} \tag{21c}$$

All the other possible anti-commutation relations yield zero.

## 3. THE NAMBU-LIKE FIELD, AND THE UNPERETURBED HAMILTONIAN AND PROPAGATORS.

The Nambu-like field of Eq. (4) may be defined by means of the elementary excitations as

$$\tilde{\Psi}_k = P_k^{-1} O_k. \tag{22}$$

The following anti-commutation relations are of interest

$$\{\tilde{\Psi}_k^+, \tilde{\Psi}_{k'}\} = \delta_{k,k'} I + \delta_{k,\bar{k}'} \alpha_0, \tag{23a}$$

$$\{\tilde{\Psi}_k, \tilde{\Psi}_{k'}\} = 0. \tag{23b}$$

The time dependence of $\tilde{\Psi}_k$ is given only by $O_k(t) = \begin{pmatrix} \gamma_k \exp(-iE_k t) \\ \eta_k \exp(-iE_k t) \\ \rho_k^+ \exp(iE_k t) \\ \sigma_k^+ \exp(iE_k t) \end{pmatrix}$, in accordance with the assumption that the four elementary excitations have a common eigenvalue. Since we wish to separate the components with different time dependence, we define

$$O_k^-(t) = \begin{pmatrix} \gamma_k \exp(-iE_k t) \\ \eta_k \exp(-iE_k t) \\ 0 \\ 0 \end{pmatrix}, \text{ and } O_k^+(t) = \begin{pmatrix} 0 \\ 0 \\ \rho_k^+ \exp(iE_k t) \\ \sigma_k^+ \exp(iE_k t) \end{pmatrix}.$$

The spatial dependence of the field should be given by a Fourier transform. Thus, the space and time dependent $\tilde{\Psi}$ is given by

$$\tilde{\Psi}(x,t) = \frac{1}{2} \sum_k P_k^{-1} [O_k^-(t) + O_k^+(t)] \exp(ikx). \tag{24}$$

The unperturbed Hamiltonian density is given by

$$H_0(x) = i\tilde{\Psi}^+(x,t) \frac{d}{dt} \tilde{\Psi}(x,t). \tag{25}$$

In the calculation of the Hamiltonian we drop the parts that are oscillating in time. We get

$$H_0(x) = \frac{1}{4}\sum_{k,k'} E_k [P_{\gamma,k'}P_{\gamma,k}^{-1}\gamma_k^+\gamma_k + P_{\eta,k'}P_{\eta,k}^{-1}\eta_k^+\eta_k - P_{\rho,k'}P_{\rho,k}^{-1}\rho_k\rho_k^+ - P_{\sigma,k'}P_{\sigma,k}^{-1}\sigma_k\sigma_k^+]$$

$$+ \frac{1}{4}\sum_{k,k'} E_k [P_{\gamma,k'}P_{\eta,k}^{-1}\gamma_k^+\eta_k + P_{\eta,k'}P_{\gamma,k}^{-1}\eta_k^+\gamma_k - P_{\rho,k'}P_{\sigma,k}^{-1}\rho_k\sigma_k^+ - P_{\sigma,k'}P_{\rho,k}^{-1}\sigma_k\rho_k^+] \qquad (26)$$

where $P_{\gamma,k}^{-1}$, and $P_{\gamma,k'}$ are the $\gamma$ column of $p_k^{-1}$, and $\gamma$ row of $P_{k'}$, respectively. Similar definitions apply for the other rows and columns of $p_k^{-1}$ and $P_{k'}$. The first sum vanishes unless k=k', while the second vanishes unless $k = \bar{k}'$. Thus, we get

$$H_0 = \frac{1}{2}\sum_k E_k [O_k^+\beta O_k + \frac{2u_k(v_k - \theta_k)}{1+2w_k} O_k^+\beta\tau_3 O_k]. \qquad (27)$$

The first term in Eq. (27) is consistent with the assumption of identical eigenvalues for the four elementary excitations, but the second term violates this assumption. Accepting the correction of the second term, besides the inconsistency, complicates considerably the unperturbed propagator. However, a fast evaluation of this term suggests that it is small and could approximately be neglected. Thus, we approximate

$$H_0 \cong \frac{1}{2}\sum_k E_k O_k^+ \beta O_k . \qquad (28)$$

When the Hamiltonian is applied to the ground state- $H_0 |\Phi_0>$, the first two terms yield zero, but the last two terms (that contain $\rho_k\rho_k^+$, and $\sigma_k\sigma_k^+$) yield $\sum_{k(\varepsilon_k<0)} -E_k |\Phi_0>$. Thus, if one wants to fix the eigenvalue of the ground state to be zero, this constant term has to be subtracted from the Hamiltonian.

Expressing the Hamiltonian by means of $\tilde{\Psi}$, and consequently by means of the $c_k$'s is useful. Eq. (18) and Eq. (28) suggest that

$$H_0 \cong \frac{1}{2}\sum_k E_k \tilde{\Psi}_k^+ (P_k^{-1}\beta P_k)\tilde{\Psi}_k = \frac{1}{2}\sum_k E_k \tilde{\Psi}_k^+ X_k \tilde{\Psi}_k . \qquad (29)$$

The matrix $X_k$ is easily calculated to be

$$X_k = -2w_k\alpha_3 - 2u_k\alpha_1 + (\theta_k - v_k)\beta , \qquad (30)$$

with $\alpha_1$, and $\alpha_3$, the known Dirac matrices. Thus, the unperturbed Hamiltonian is

$$H_0 = \frac{1}{2}\sum_k [\varepsilon_k \tilde{\Psi}_k^+ \beta \tilde{\Psi}_k + \Lambda_k \tilde{\Psi}_k^+(x)\alpha_3 \tilde{\Psi}_k(x) + \Delta_k \tilde{\Psi}_k^+ \alpha_1 \tilde{\Psi}_k]. \tag{31}$$

Notice that the sum is over all momenta in the Brillouin zone but not over the spin states, because the two spin states are already included within each $\tilde{\Psi}_k$. Due to the structure of $\tilde{\Psi}_k$, each energy is counted four times, two times are compensated by not summing over the spin states, and the other two by dividing the sum by 2. Examining of the unperturbed Hamiltonian suggests immediately that the energy of the un-condensed state, the pseudo gap, and the superconductive gap are given by

$$\varepsilon_k = E_k (\theta_k - v_k) \tag{32a}$$

$$\Lambda_k = -2 w_k E_k \tag{32b}$$

$$\Delta_k = -2 u_k E_k \tag{32c}$$

Which suggest immediately that $E_k^2 = \varepsilon_k^2 + \Lambda_k^2 + \Delta_k^2$. With these parameters, the second term of Eq. (27), which has been assumed small and dropped in Eq. (28) is zero at the Fermi level. It is also insignificantly small at high energies $\varepsilon_k \gg \Lambda_k$. It gets to its maximum value around $\varepsilon_k = \Lambda_k$, where it becomes $\frac{\Delta_k \Lambda_k}{E_k + \Lambda_k} \approx \frac{\Delta_k}{(\sqrt{2}+1)} \ll E_k$. This justifies the approximation we made in Eq. (28).

Notice that the spatial dependence in Eq. (31) is denoted only for the $\alpha_3$ term, because the others are independent upon x. The spatial dependence stems from terms with $c_{\bar{k}}^+$ and $c_{-\bar{k}}^+$ in Eq. (24) which are multiplied by $\exp(ikx)$, rather than by $\exp(i\bar{k}x)$. Fortunately, the difference is constant (for each nested section in the Fermi surface), so that the sum could anyway be carried out. The explicit spatial dependence of the $\alpha_3$ term in Eq. (31) is

$$\frac{1}{2}\Lambda_k \tilde{\Psi}_k^+(x)\alpha_3 \tilde{\Psi}_k(x) =$$

$$= \frac{1}{2}\Lambda_k [(c_{\bar{k},s}^+ c_{k,s} + c_{-k,-s}^+ c_{-\bar{k},-s})\exp i(\mathbf{k}-\bar{\mathbf{k}})\cdot \mathbf{x} + (c_{-\bar{k},-s}^+ c_{-k,-s} + c_{k,s}^+ c_{\bar{k},s})\exp i(\bar{\mathbf{k}}-\mathbf{k})\cdot \mathbf{x}], \tag{33}$$

which is in agreement with Ref. [3]. Notice also that an equivalent spatial dependence should exist in the ground state of Eq. (5). Although the ground state is a combination of products of the above exponents, it results in terms with the above exponents (plus constant terms), when reduced to the first Brillouin zone. This spatial dependence was discussed in [3] to produce charge or spin density wave. It is responsible for the violation of the momentum conservation in electron scatterings.

The anti-commutation relations of Eqs. (21) and (23) suggest the existence of the propagators $G_0(\bar{k},k,\omega)$, and $G_0(k,\bar{k},\omega)$, in addition to the ordinary propagator $G_0(k,\omega)$. Let us start by calculating $G_0(k,\omega)$

$$G_0(k,t) = -i <\Phi_0 | T\{\tilde{\Psi}_k(t), \tilde{\Psi}_k^+(0)\}|\Phi_0>, \qquad (34)$$

where $T$ is the time ordering operator. The calculation proceeds as usual [3,17], and we get

$$G_0(k,t) = -i <\Phi_0 | M_k \exp(-iE_k t)\Theta(t) - N_k \exp(iE_k t)\Theta(-t)|\Phi_0>, \qquad (35)$$

where $\Theta$ is the known step function, and

$$M_k = P_{\gamma,k}^{-1} P_{\gamma,k} + P_{\eta,k}^{-1} P_{\eta,k} = \frac{1}{2}[I - 2w_k \alpha_3 - 2u_k \alpha_1 + (\theta_k - v_k)\beta], \qquad (36a)$$

$$N_k = P_{\rho,k} P_{\rho,k}^{-1} + P_{\sigma,k} P_{\sigma,k}^{-1} = \frac{1}{2}[I + 2w_k \alpha_3 + 2u_k \alpha_1 - (\theta_k - v_k)\beta], \qquad (36b)$$

where $P_{\gamma,k}^{-1}$, $P_{\gamma,k}$, $P_{\eta,k}^{-1}$, $P_{\eta,k}$ have been defined just after Eq. (26). The same notation applies for $\rho$ and $\sigma$, apart from the exchange of the matrices $P_k$ and $P_k^{-1}$, due to the time ordering operator. Eq. (35) is Fourier transformed to obtain

$$G_0(k,\omega) = \frac{M_k}{\omega - E_k + i\delta} + \frac{N_k}{\omega + E_k - i\delta} = \frac{\omega I + \Lambda_k \alpha_3 + \Delta_k \alpha_1 + \varepsilon_k \beta}{\omega^2 - E_k^2 + i\delta}. \qquad (37)$$

The propagator of Eq. (37) is of the expected form, and it is in agreement with [2]. The non-diagonal propagator $G_0(\bar{k},k,t)$ is given by

$$G_0(\bar{k},k,t) = -i <\Phi_0 | T\{\tilde{\Psi}_{\bar{k}}(t), \tilde{\Psi}_k^+(0)\}|\Phi_0>$$

$$= -i <\Phi_0 | M_{\bar{k}k} \exp(-E_k t)\Theta(t) - N_{\bar{k}k} \exp(iE_k t)\Theta(-t)|\Phi_0>, \qquad (38)$$

which is Fourier transformed to give

$$G_0(\bar{k},k,\omega) = \frac{M_{\bar{k}k}}{\omega - E_k + i\delta} + \frac{N_{\bar{k}k}}{\omega + E_k - i\delta}. \tag{39}$$

A straight forward calculation yields $M_{\bar{k}k} = \alpha_0 M_k$, and $N_{\bar{k}k} = \alpha_0 N_k$, which yields

$$G_0(\bar{k},k,\omega) = \alpha_0 G_0(k,\omega). \tag{40a}$$

In the same manner, one gets

$$G_0(k,\bar{k},\omega) = G_0(k,\omega)\alpha_0, \tag{40b}$$

$$G_0(\bar{k},\bar{k},\omega) = \alpha_0 G_0(k,\omega)\alpha_0. \tag{40c}$$

Eqs. (40) are important, because they suggest the convenient use of only momentum defined propagators in the perturbation theory, provided that the multiplying matrices $\alpha_0$ are associated with interaction vertices [3].

4. **THE INTERACTION HAMILTONIAN, AND PERTURBATION THEORY WITH NON-DIAGONAL VERTICES.**

The interaction Hamiltonian is given by

$$H_i = \frac{1}{8}\sum_{k,k',q} V_q (\tilde{\Psi}^+_{k'-q}\tau_3\tilde{\Psi}_{k'})(\tilde{\Psi}^+_{k+q}\tau_3\tilde{\Psi}_k) - \frac{1}{2}\sum_k (\Lambda_k \tilde{\Psi}^+_k \alpha_3 \tilde{\Psi}_k + \Delta_k \tilde{\Psi}^+_k \alpha_1 \tilde{\Psi}_k). \tag{41}$$

Here $k$ and $k'$ are summed over the whole Brillouin zone. The pre-factor $\frac{1}{8}$ takes into account the 4-dimentional structure of the field, and that there is no sum over spin states, contrary to the notations in [1,3]. It also takes into account the double counting which stems from the symmetry between the $k$ and $k'$ in the first term. The second term compensates for the addition of the same term in $H_0$. The perturbation approximation within the Hartree-Fock scheme yields zero, since it is already assumed in $H_0$. Therefore, the calculation of the order parameters $\Lambda$ and $\Delta$, by means of field perturbation theory, should apply only the first term of Eq. (41), which hereafter is denoted by $H_i'$. Thus, $H_i'$ may be rewritten as

$$H_i' = \frac{1}{8}\sum_{k,k',q} V_q^t (\tilde{\Psi}^+_{k'-q}\tau_3\tilde{\Psi}_{k'})(\tilde{\Psi}^+_{k+q}\tau_3\tilde{\Psi}_k). \tag{42}$$

Let us now deal with the interaction potential in Eq. (42). It was found in [3] that the el-phonon-el interaction is essential, and it must be added to the bare Coulomb interaction. Essentially, this should be done by defining the field of the vibrating ions, and by incorporating into $H_i'$-the interaction of the electrons with that field. Then, contractions within the ionic fields create phonon propagators, factored by the el-phonon- interaction. Here, we make a short cut by assuming that we have already obtained the phonon propagators, and define

$$V_q^t = V_q + \sum_\lambda |g_{q,\lambda}|^2 D_{q,\lambda} , \tag{43a}$$

$$V_{\bar{q}}^t = V_{\bar{q}} + \sum_\lambda |g_{\bar{q},\lambda}|^2 D_{\bar{q},\lambda} , \tag{43b}$$

where $V_q$ is the bare Coulomb interaction, $g_{q,\lambda}$ is the bare electron-phonon matrix element, $D_{q,\lambda}$ is the bare phonon propagator, and $\bar{\mathbf{q}} = \mathbf{q} - (\mathbf{k}_\perp / |\mathbf{k}_\perp|) 2k_F$. The phonon mode is $\lambda$. Field perturbation, according to Wick's theorem, involves products of $H_i'$'s, in all orders. The fields, of the various $H_i'$'s, are "contracted" with each other. All contraction combinations are allowed, and each different combination corresponds to a different Feynman diagram. Some of these combinations are also responsible for screening the interaction. The treatment here is consistent with the treatment in [3], where we assumed regular lattice periodicity. The other relevant periodicity is the one in which the lattice was re-constructed in accordance with the "anti-ferromagnetic" order, which makes the Brillouin zone to coincide with the Fermi zone. In the present stage of the analysis, it is uncertain whether both periodicities should produce the same results. We assumed the "regular periodicity" not because it is dominant in experiment, but merely for the sake of convenience. Experiments suggest that the two kinds of periodicities exist in various HTSC, and each kind deserves investigation.

In the present investigation, as in previous ones, we assume that the Hartree integral is the dominant one in calculating the two order parameters- $\Delta$, and $\Lambda$. Hartree diagrams are usually associated with potentials with $|\mathbf{q}| = 0, \omega = 0$. Here one has also a diagram that is associated with potential of $|\bar{\mathbf{q}}| = 2k_F, \omega = 0$. The first was shown to vanish due to the mutual cancellation of the Coulomb interaction, and the el-phonon-el interaction [17]. The second is therefore the mere contributor. Its screened potential-$U_H$, which is terminated by two $\alpha_3$ vertices, is given by [3]

$$\alpha_3 U_H \alpha_3 = \alpha_3 V_{\bar{q}}^t \alpha_3 [1 + (V_q^t + V_{\bar{q}}^t)(\Pi_q + \Pi_{\bar{q}})]^{-1} .$$

$$\xrightarrow[q\to 0]{} \alpha_3 V_{\bar{q}}^t \alpha_3 (1 + V_{\bar{q}}^t \Pi_{\bar{q}})^{-1}. \tag{44}$$

The quantities $\Pi_q$, and $\Pi_{\bar{q}}$ are irreducible polarizations, and one can show that $\Pi_q \xrightarrow[q\to 0]{} 0$. Although $V_{q=0}^t = 0$, $V_{\bar{q}}^t$ is finite and negative, and $U_H$ was evaluated to be negative and large. The potential of Eq. (44) was obtained for HTSC with correlation pseudogap, but in the normal state. Its validity is approximately assumed also in the present analysis, in the SC state, provided that $E_k^2 = \varepsilon_k^2 + \Lambda_k^2$ is replaced by $E_k^2 = \varepsilon_k^2 + \Lambda_k^2 + \Delta_k^2$.

Eq. (44) guarantee the existence of a finite Hartree diagram for the order parameter $\Lambda$, provided that an interaction vertex is spanned by $\alpha_3$. To show this we note that $\tilde{\Psi}_k^+ \alpha_0 = \tilde{\Psi}_k^+$, and by redefining the dummy indices k and k' in Eq. (42), part of the sum could be rewritten to apply to Hartree diagrams

$$H_i'(Hartree) = \frac{1}{8} \sum_{k,k',q\to 0} V_{\bar{q}}^t (\tilde{\Psi}_{k'-q}^+ \alpha_3 \tilde{\Psi}_{k'})(\tilde{\Psi}_{k+q}^+ \alpha_3 \tilde{\Psi}_k). \tag{45}$$

In Eq. (45), $|\bar{q}| = 2k_F$, which implies that **k** is scattered into $\bar{\mathbf{k}}$, causing **k'** from the opposite nested section to scatter into $\bar{\mathbf{k}}'$.

Eq. (45) implies immediately the existence of a finite Hartree integral for $\Lambda$, but not for $\Delta$. The large experimental $\Delta$'s that have been obtained in most HTSC are not likely to result only from Fock integrals. They are more likely to be enhanced by Hartree integrals. The feasibility of a finite Hartree diagram for $\Delta$ should be sought in association with another field- a coherence field [17]. We define this coherence field- $\Psi_k$ by

$$\Psi_k = X_k \tilde{\Psi}_k = P_k^{-1} \beta O_k, \tag{46}$$

where $X_k$ is given by Eq. (30). Notice that $X_k$ is Hermitian and Unitary- $X_k = X_k^T = X_k^{-1}$. Therefore, the former results about $H_0$ and $G_0(k,\omega)$, are still valid with the coherence field. The sum of the interaction Hamiltonian that implies to Hartree diagrams is

$$H_i' = \frac{1}{8} \sum_{k,k',q\to 0} V_{\bar{q}}^t [(\Psi_{k'-q}^+ X_{k'-q}^{-1} \alpha_3 X_{k'} \Psi_{k'})(\Psi_{k+q}^+ X_{k+q}^{-1} \alpha_3 X_k \Psi_k)]. \tag{47}$$

A direct calculation gives

$$X_k^{-1}\alpha_3 X_k = (-1+8w_k^2)\alpha_3 + 8w_k u_k \alpha_1 - 4w_k(\theta_k - v_k)\beta, \tag{48a}$$

$$= E_k^{-2}[(-E_k^2 + 2\Lambda_k^2)\alpha_3 + 2\Lambda_k \Delta_k \alpha_1 + 2\Lambda_k \varepsilon_k \beta]. \tag{48b}$$

This group includes the vertices $\alpha_1, \alpha_3$, and $\beta$, whose self-energies are obtained by both Hartree and Fock integrals. The self-energy for $I$, which yields the renormalization of the frequency parameter, is obtained only by a Fock integral. The pre-factors in Eqs. (48) are coherence parameters. As such they are probability amplitudes, whose squares sum to unity, $(-1+8w_k^2)^2 + (8w_k u_k)^2 + 16w_k^2(\theta_k - v_k)^2 = 1$. Without nesting the vertices of Eq. (47) are not relevant, $w_k = 0$, and one is left only with the vertices $X_{k+q}^{-1}\tau_3 X_k \xrightarrow{q \to 0} (1-8u_k^2)\tau_3 + i\alpha_2\beta 4u_k(\theta_k - v_k)$, of Eq. (42). Moreover, since the $\bar{k}$ indices lose meaning, 4-dimensional matrices degenerate into 2-dimensional, and the vertex $i\alpha_2\beta$ degenerates to the two dimensional $\tau_1$ vertex. Then the vertices of Eq. (42) turn into $X_k^{-1}\tau_3 X_k = E_k^{-2}[(\varepsilon_k^2 - \Delta_k^2)\tau_3 + 2\Delta_k \varepsilon_k \tau_1]$, which is consistent with Ref. [17]. Alternatively, when superconductivity is turned off in HTSC with correlation gap, $u_k = 0$, and one gets $X_k^{-1}\alpha_3 X_k = -E_k^{-2}[(\varepsilon_k^2 - \Lambda_k^2)\alpha_3 - 2\Lambda_k \varepsilon_k \beta]$, which is consistent with Eq. (3c). Thus, the field of Eq. (45), and the vertices of Eqs. (48) are fully consistent with the concept of coherence fields in former analysis [3,17].

The Hartree diagram for the self-energy is shown in Fig. 1. It is given by

$$\Sigma_H = -iX_k^{-1}\alpha_3 X_k U_H(q=2k_F, \omega=0)\frac{1}{8}Tr\int \frac{d\nu d^3k'}{(2\pi)^4} X_{k'}^{-1}\alpha_3 X_{k'} G(k',\nu)\exp(i\delta\nu). \tag{49}$$

Eq. (49) gives the three components of the Hartree self-energy- the $\alpha_1, \alpha_3$, and the $\beta$ components. The components have a common integral, and differ from each other only through their coherence factors. If in Eq. (49) we approximated the Green's function- $G(k',\nu)$ by its unperturbed counterpart- $G_0(k',\nu)$, the integrand may be simplified, and after the frequency integration, we get

$$\Sigma_H = -X_k^{-1}\alpha_3 X_k U_H(q=2k_F, \omega=0)\int \frac{d^3k'}{(2\pi)^3}\frac{\Lambda_{k'}}{4E_{k'}}. \tag{50}$$

There are two differences between Eq. (50), and Eq. (9) of [2]. One is the coherence factor $X_k^{-1}\alpha_3 X_k$, and the other one is the division by 4. The integral may be transformed approximately into $\int_{-\Lambda}^{\Lambda} d\varepsilon_{k'} N_{nes}\frac{\Lambda_{k'}}{4\sqrt{\varepsilon_{k'}^2 + \Lambda_{k'}^2 + \Delta_{k'}^2}} \approx aN_{nes}\Lambda$, where $N_{nes}$ is the density of states of the un-condensed phase by the two nesting sections of the states $k$ and $\bar{k}$. $\Lambda$ is the value of $\Lambda_k$ at the Fermi level, and $a$ is a positive numerical factor, $a \approx 0.5$. Thus, we have

$$\Sigma_H = -U_H(\omega=0, q=2k_F)aN_{nes}\Lambda E_k^{-2}[(-E_k^2 + 2\Lambda_k^2)\alpha_3 + 2\Lambda_k\Delta_k\alpha_1 + 2\Lambda_k\varepsilon_k\beta]. \tag{51}$$

The three components of the Hartree self-energy are given approximately by

$$\Lambda_{k,H} = -U_H aN_{nes}\Lambda \frac{\Lambda_k^2 - \Delta_k^2 - \varepsilon_k^2}{\Lambda_k^2 + \Delta_k^2 + \varepsilon_k^2}, \tag{52a}$$

$$\Delta_{k,H} = -U_H aN_{nes}\Lambda \frac{2\Lambda_k\Delta_k}{\Lambda_k^2 + \Delta_k^2 + \varepsilon_k^2}, \tag{52b}$$

$$\chi_{k,H} = -U_H aN_{nes}\Lambda \frac{2\Lambda_k\varepsilon_k}{\Lambda_k^2 + \Delta_k^2 + \varepsilon_k^2}. \tag{52c}$$

The quantity $\chi_k$ acts to renormalize the energy according to: $\tilde{\varepsilon}_k = \varepsilon_k + \chi_k$. Notice that since $U_H$ is negative, the signs of $\chi_k$, and $\varepsilon_k$ are identical. Consequently, the eigenvalue of the ground state with the coherence field is lower than with the Nambu-like field, an outcome which suggests the validity of the coherence field. At very high energies $\chi_k$ decays to zero, and the renormalization factor decays to unity. Eq. (52a) suggests that at zero energy, $\Lambda_{0,H} \approx -U_H aN_{nes}\Lambda$, that it changes sign at $\varepsilon_k^2 = \Lambda_k^2 - \Delta_k^2$, and becomes $\Lambda_{\infty,H} \approx U_H aN_{nes}\Lambda$ at very large energies. Eq. (52b) suggests that $\Delta_{k,H}$ is roughly constant for energies smaller than $\Lambda_k$, and it decays to zero (as $\varepsilon_k^{-2}$) at much higher energies. Note that the quantities on the left hand sides of Eqs. (52) are Hartree values, while their counterparts on the right hand sides are the sums of the Hartree and the Fock values. As in [1,2], we assume that $\Lambda_H = \Lambda(1-f_\Lambda)$, $\Delta_H = \Delta(1-f_\Delta)$, and that $f_\Lambda, f_\Delta < 0.5$. Eqs. (52) imply immediately that $2f_\Lambda - f_\Delta \approx 1$. Assuming positive values for $f_\Lambda$ and $f_\Delta$ contradicts the assumption that both Fock integrals are small relative to their Hartree counterparts. We conjecture, therefore, that $f_\Delta$ is negative, so that one should have $2|f_\Lambda| + |f_\Delta| \approx 1$. This last evaluation, though, is conditioned upon the accuracy of the approximations we have made.

The Fock integral is

$$\Sigma_F(k,\omega) = i4X_k^{-1}\tau_3 \int \frac{dvd^3k'}{(2\pi)^4} X_k G(k',v) U_F(\mathbf{q}=\mathbf{k}-\mathbf{k}',\omega-v)X_{k'}^{-1}\tau_3 X_k. \tag{53}$$

Here $U_F(q,\nu)$ is the momentum and frequency dependent screened potential that is appropriate to Fock integrals. The Fock integral of Eq. (53) is hard to evaluate for the following reasons: 1) Contrary to the Fock integral for ordinary superconductivity, which yields the known Eliashberg equation, here one cannot assume the static limit for the Coulomb part of the potential. The structure of the electron propagator suggests that the screened potential might have some spectral "structure" around energies of the order of $\Lambda$, which is also the energy scale for the variations of the electronic propagator. 2) Additional difficulty is introduced by the coherence factor matrix. For these reasons, I would rather leave Eq. (53) as is. The analyses of [1] and [2] were not based on the coherence field, and did not involve coherence factors. If one is to speculate by using these conditions, and also by ignoring difficulty # 1, then one may show that $f_\Delta$ is indeed negative, and $|f_\Delta|$, and $|f_\Lambda|$ are small.

The former discussion suggests that the analysis is still incomplete. It lacks mainly the Fock part, and the incorporation of disorder. The innovation of the present analysis is the two components of the Hartree off-diagonal self energy. Eqs. (52) give only the relative size of these Hartree parameters. At zero energy these should be $1 - f_\Lambda = -U_H N_{nes} a$, and $1 - f_\Delta = -U_H N_{nes} 2a$. We assume that the interaction should be substantial, $|U_H N_{nes} a| \geq 0.5$. Otherwise, the theory is quantitatively insignificant, even in cases where Hartree self- energies are finite. This, in turn, requires that the el-phonon-el interaction for wave-vectors $|\mathbf{q}| = 2K_F$, in directions normal to nesting surfaces, must be significantly large.

5. CONCLUDING REMARKS.

We have established the double correlated phase and its treatment by means of field perturbation. The present paper is a step in establishing a theory of HTSC, which is based on the nesting feature of the Fermi surfaces of these materials. We have shown that Hartree integrals are essential for obtaining large order parameters, and we have established their validity. The reader has probably noticed that the whole analysis of the present paper deals with insolating materials with both correlation and superconductive order parameters. This does not mean that we suggest that an insolating material may become superconductive. The analysis has been performed this way only as a matter of convenience. The way of matching the present analysis to actual conducting HTSC may be deduced from [1], and [2], by simply applying the transformation $\omega \rightarrow \omega + i\Gamma sign(\omega)$, where $\Gamma$ is a quasi-particle scattering rate due to disorder. We have seen that such a transformation produces states in the correlation gap, and provides electron conductivity to the material. Further development, beyond the perception in [1] and [2], might suggest equivalent disorder induced transformations to $\Lambda$, and to $\Delta$.

Another idealization that has been done in the present analysis is the assumption of 100% nesting states. Adopting a more realistic situation may be deduced from [2], where it was assumed that only a fraction of the states by the Fermi level are nested- $N_0^{nes}$, where the other part is not- $N_0^{n.n}$. Both the superconductive and the correlation parameters are obtained from integrals, and the large values that obtained from the nested sections should enhance the values of the other sections.

Having made the above remark to match the analysis to conductors, we again assume a perfectly nested insolating material. Although such a material cannot exhibit infinite conductivity and the Meissner effect, its quasi-particles may have double correlations, with one type that is superconductive-like (its part in the ground state does not preserve the particle number). This is an essentially interesting feature, the full consequences of which are still unknown. One such a consequence though is clear, it enhances the gap, from $\Lambda$ to $\sqrt{\Lambda^2 + \Delta^2}$. A critical experimental test to the existence of the double correlated phase in insolating un-doped mother HTSC may be a proximity experiment. If such a material with a well characterized surface would be covered with a thin metallic film, which would exhibit an enhanced superconductivity, this should prove the existence of the double correlated phase in the insulator.

REFERENCES.

1. Moshe Dayan, J. Supercon. **17**, 487 (2004).
2. Moshe Dayan, J. Supercon. **17**, 739 (2004).
3. Moshe Dayan, J. Supercon. **20**, 239 (2007). See also erratum in J. Supercon. **20**, 341 (2007).
4. (a) Y. Nambu, Phys. Rev. 117, 648 (1960). (b) L. P. Gorkov, JEPT **7**, 505 (1958).
5. Moshe Dayan, J. Supercon. **17**, 353 (2004).
6. S. K. Chan, and V. Heine, J. Phys. F. **3**, 795 (1973).
7. D. J. Scalapino, E. Loh, Jr. J. E. Hirsch, Phys. Rev. **B35**, 6694 (1987).
8. L. F. Mattheiss, Phys. Rev. Let. **58**, 1028 (1987).
9. J. Yu, A. J. Freeman, and J. H. Xu, Phys. Rev. Let. **58**, 1035 (1987).
10. L. F. Mattheiss, and D. R. Hamann, Phys. Rev. Let. **60**, 2681 (1988).
11. S. Mssidda, J. Yu, and A. J. Freeman, Physica C, **152**, 251 (1988).
12. D. Vaknin, S. K. Sinha, D. E. Moncton, D. C. Johnston, J. M. Newsman, C. R. Safinya, and H. E. King, Jr. Phys. Rev. Let. **58**, 2802 (1987).
13. Y. J. Uemura, W. J. Kossler, X. H. Yu, J. R. Kempton, H. E. Schone, D. Opie, C. E. Stronach, D. C. Johnston, M. S. Alvarez, and D. P. Goshorn, Phys. Rev. Let. **59**, 1045 (1987).
14. P. Aebi et al. Phys. Rev. Let. 72, 2757 (1994).
15. A. A. Kordyuk et al. Phys. Rev. B 66, 14502 (2002).
16. Moshe Dayan, in Models and Methods of High-$T_c$ Superconductivity, J. K. Srivastava and S. M. Rao, eds. (Nova Scientific, New York, 2003), p-160-166, and Refs. therein.
17. Moshe Dayan, J. Supercon. **19**, 477 (2006).

FIGURE CAPTIONS

Fig. 1

The diagram for the Hartree off-diagonal self-energy. The diagram has three components- $\alpha_1, \alpha_3$, and $\beta$. Each component is weighted by the appropriate coherence parameter, and calculated as a sum of three integrals. Each integral yields a scalar since it is obtained by the trace of the product of two identical Dirac matrices, the vertex and the off-diagonal component of the ring propagator. Note that the notation of two separate sums of $(\alpha_1 + \alpha_3 + \beta)$, one for the vertices and the other for the propagators, is only symbolic. The matrix notation should actually be: $Tr\{\int (\alpha_1 C_1 + \alpha_3 C_3 + \beta C_\beta)(\alpha_1 G_1 + \alpha_3 G_3 + \beta G_\beta)\}$, where the $C$'s are the coherence parameters, and the $G$'s are the propagator components.

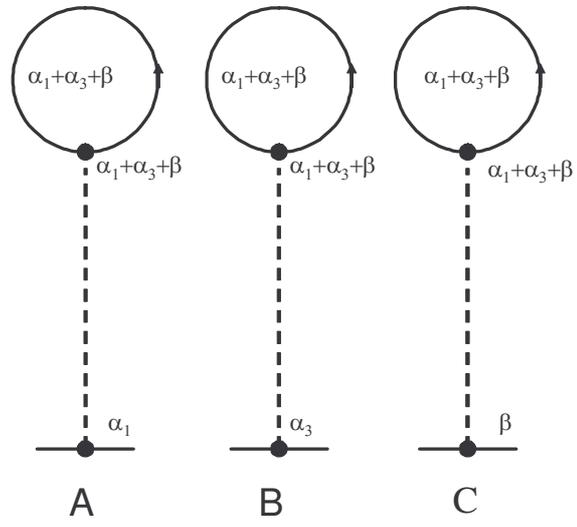